# SYNCHRONIZATION OF THE FERMILAB BOOSTER AND MAIN INJECTOR FOR MULTIPLE BATCH INJECTION

R. Zwaska*, S. Kopp, University of Texas at Austin, Austin, TX, 78712, USA
W. Pellico#, R. Webber, Fermi National Accelerator Laboratory, Batavia, IL 60510, USA

*Abstract*
  To date, the 120 GeV Fermilab Main Injector accelerator has accelerated a single batch of protons from the 8 GeV rapid-cycling Booster synchrotron for production of antiprotons for Run II. In the future, the Main Injector must accelerate 6 or more Booster batches simultaneously; the first will be extracted to the antiproton source, while the remaining are extracted for the NuMI/MINOS (Neutrinos at the Main Injector / Main Injector Neutrino Oscillation Search) neutrino experiment. Performing this multi-batch operation while avoiding unacceptable radioactivation of the beamlines requires a previously unnecessary synchronization between the accelerators. We describe a mechanism and present results of advancing or retarding the longitudinal progress of the Booster beam by active feedback radial manipulation of the beam during the acceleration period.

## INTRODUCTION

The Booster accelerates protons from 400 MeV, kinetic energy, to 8 GeV. The Main Injector (MI) accepts the 8 GeV protons and accelerates them further to 120 GeV. The MI circumference is seven times that of the Booster and can accept several batches of protons from the Booster. The forthcoming programs of the NuMI [1] neutrino beam and slip-stacking for anti-proton production [2] will require this multiple-batch operation continuously.

The Booster's ability to deliver the necessary protons to these programs is limited by the radio-activation of its components caused by uncontrolled beam loss. A significant portion of this loss occurs at extraction. The Booster beam is extracted in a single turn by the firing of a series of kicker magnets displacing the beam into the field region of a septum magnet. However, the risetime of the kicker magnets is 30-40 ns, while the bunch spacing is 19 ns. Thus, one or two bunches of protons would be lost on the septum.

The remedy to this loss is to remove the charge from a number of buckets (currently three) at the start of the Booster cycle. This "notch" is created by the fast firing of a single kicker magnet, clearing out the buckets; this has already been implemented to service single-batch applications of the Booster.

Proper notching requires the beam extraction to be synchronized with the notch in the beam. However, multi-batch operation will require the extraction to also be synchronized with the already circulating beam in the MI.

___________________________
*zwaska@mail.hep.utexas.edu
#pellico@fnal.gov

As will be discussed, the Booster beam is *a priori* not synchronized with the MI. Furthermore, the Booster undergoes cycle-to-cycle variations in the amount of longitudinal slippage experience by beam in the Booster relative to the MI.

Providing the necessary synchronization (dubbed "cogging") requires active feedback during the acceleration cycle since the Booster has no flattop where the beam can be manipulated longitudinally at leisure, cf. [3]. As the MI frequency is fixed while awaiting injection, the process of cogging occurs entirely within the Booster. We present an analysis of the problem as well as a working system to provide the necessary cogging. This works follows an earlier investigation that established the system of measurement and provided a proof-of-principle [4].

## THE BOOSTER

The Booster is a rapid-cycling synchrotron with combined function magnets. It is provided 400 MeV, debunched 200 MHz H⁻ ions from a linear accelerator. The H⁻ ions are passed through a stripping foil at injection to produce protons, filling the entire ring.

The Booster magnets are part of a 15 Hz resonant circuit that is locked to the wall-socket frequency. As discussed in the next section, variation of this frequency is a source of error. The beam momentum, to first order, can be approximated as a sinusoid:

$$p(t) = p_0 - p_1 \cos(2\pi f t) \qquad (1)$$

Where $p_0 = (p_e + p_i)/2 = 4.9223$ GeV/$c$ and $p_1 = (p_e - p_i)/2 = 3.9666$ GeV/$c$; $p_i$ and $p_e$ are the Booster injection and extraction momenta.

The debunched beam is adiabatically captured into 84, 37.77 MHz RF (radio frequency) buckets. To accommodate the sinusoidal ramp capture is accomplished quickly (< 1 ms).

Orbit feedback is initiated shortly after the bunched structure emerges. Radial position is measured by a BPM (Beam Position Monitor) in a low dispersion region (~2m). That position is compared to a preprogrammed position which varies during the cycle. The phase of the RF with respect to the beam is adjusted to force the radial position toward the programmed position. The feedback is inverted after transition which occurs at $\gamma_t = 5.45$, approximately 18 ms into the cycle.

The Booster frequency changes substantially during the cycle, ramping to a final value of 52.8114 MHz in 33 ms. The frequency follows a rough frequency curve, fine-tuned by the feedback maintaining radial position. The final frequency and phase matching to the MI RF is accomplished by a phase-lock module that applies radial feedback in the last 3 ms. This module corrects the phase of the Booster RF to that of the MI to within a few degrees. The process is similar to cogging except for its magnitude, as it corrects differences of 1 bucket, while cogging must be able to correct for the entire 84 in the ring.

## SOURCES OF SLIPPAGE

The Booster frequency at injection is substantially lower than the MI's. As such, the Booster beam will slip substantially with respect to the circulating beam in the MI, a total of about 100,000 buckets. Any variation in the Booster frequency during the cycle will result in a difference at the end. (There are several methods to analyze this slippage, the relative frequency was found to be the most convenient).

We can define a relative slippage rate η:

$$\eta(t) = f_{MI} - f_B(t) \quad (2)$$

and a total slippage for any time in the cycle:

$$S(t) = \int_0^t dt' \, \eta(t') \quad (3)$$

where $f_{MI}$ is the fixed MI frequency of 52.8114 MHz and $f_B$ is the Booster frequency that varies throughout the cycle.

Variations in slippage can them be parameterized as a function of time for any variation that would result in the Booster frequency changing in time. We will consider four types of variation, all of which are related to the regulation of the magnet resonant circuit.

The wallsocket line frequency is known to vary by as much as 30 mHz over a few minutes. This would result in an 8 mHz variation in the Booster magnet frequency, corresponding to a 9 μs difference in cycle time. Furthermore, as the injection time is chosen by extrapolation from the previous cycles, a similar timing error can be introduced at the start of the cycle. Additionally, the magnet currents are known to vary at the part-per-thousand level, so variations in $p_i$ and $p_e$ were considered.

Variation in slippage due to an error in an arbitrary variable, $\delta x$, is calculated to first order as:

$$\delta S(t) = \delta x \int_0^t dt' \, \frac{d\eta}{dx}(t') \quad (4)$$

Each of the four errors considered have a different evolution in time. The shapes of these curves are shown in figure 1.

The magnitudes of the errors were also calculated: a 1 μs timing error results in a total slippage of 15 buckets; 1 mHz frequency error results in a slippage of 7 buckets (separate from the ensuing timing error); a 1/10,000 variation of injection magnet current results in a 10 bucket slippage; and a 1/10,000 variation of extraction magnet current would result in a 7 bucket slippage. (It should be noted that the slippage due to magnet current variations are overstated as the injection frequency is imposed on the beam, before feedback is enabled; and the extraction frequency is imposed on the beam as phase-lock begins.)

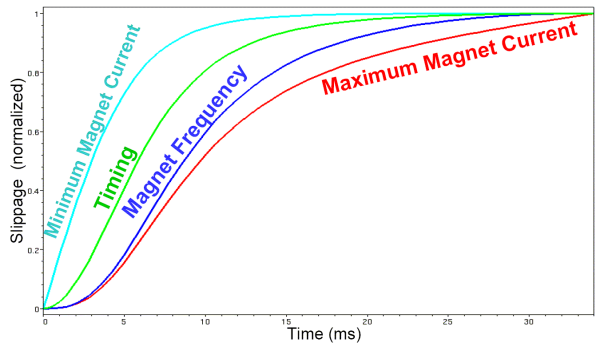

Figure 1: Time evolution of relative slippage for four possible source of cycle-to-cycle variation.

## SYNCHRONIZATION METHOD

*Beam Measurement System*

A specific system was constructed to monitor the position of the notch, or an arbitrary revolution marker through out the Booster cycle. This system is predominantly that described in [4].

A set of 3000 measurements (one at each MI revolution) is made during the Booster cycle, giving a trajectory of the Booster with respect to the MI. As noted previously, the Booster frequency is initially much less than that of the MI, so the slippage is large. However, the slippage can be compared to that of a previous cycle, giving a relative slippage. Several such relative trajectories are shown in figure 2. The cycle-to-cycle variation is as large as three revolutions of the Booster.

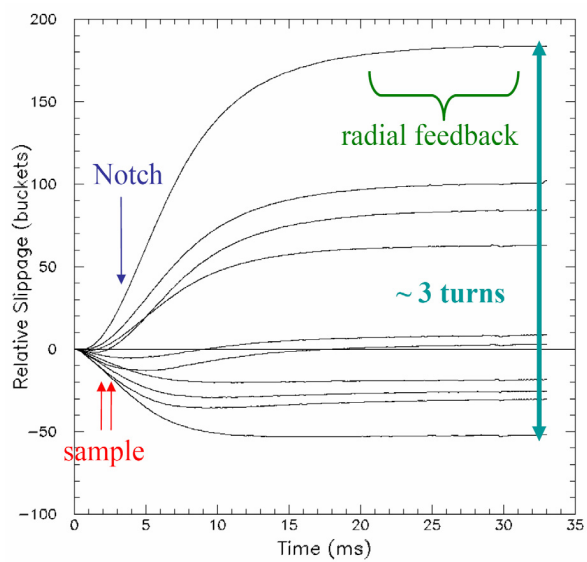

Figure 2: Typical slippage relative to a baseline cycle – a range of as much as three revolutions. Shown also are the regions used to correct for the slippage.

This system was found to have an intrinsic timing error due to its use of the MI revolution marker as a trigger. The marker comes at an arbitrary time with respect to the Booster injection, resulting in a variation of as much as 11 μs, the MI revolution period. This error was eliminated by keeping an internal marker synchronized to the Booster magnet cycle. Subsequently, the range of relative slippage was reduced by 40 %.

*Feedback System*

The first few ms of the cycle give some indication of the ultimate slippage that will occur. The creation of the notch is delayed several ms so this initial measurement can be made and the ultimate slippage estimated. The notch is then made anticipating this slippage.

A delay of 5 ms is about the maximum delay possible to still make a complete notch, as the beam grows stiffer through the cycle. Additionally, the notching losses are marginally higher due to the increased energy. These considerations limit how late the notch can be made.

The prediction is made by performing a fit to the first few hundred measurements and extrapolating to the end of the cycle. Due to the limited predictive power, a two parameter fit was found to be adequate. The signal to the notcher is made at the appropriate time to compensate for the predicted slippage.

The remaining error is corrected later in the cycle by radial feedback. After transition, a signal is generated, corresponding to the size of the correction needed; it is applied to the radial feedback loop as described earlier. This is an offset in the preprogrammed radial position curve. This results in an induced slippage of approximately $d\eta/dr = 0.5$ buckets / ms / mm.

This feedback can also be applied before transition, but is more difficult for two reasons: the ultimate slippage is not reached until after transition, so the correction cannot be complete; and the beam is much larger, so only a small displacement is possible without intersecting the beam pipe.

Experiments after transition showed that displacements less than 8 mm (at the position of the radial feedback BPM) resulted in no loss. However, increases in beam emittance were observed for displacement greater than 4 mm.

## SYNCHRONIZATION RESULTS

A series of cogged trajectories is shown in figure 3. The notch prediction reduces the spread to about 25 buckets (except for a few outliers). The radial feedback further reduces the spread to a ± 1 bucket error at the end of the cycle. The radial displacement was about 4 mm here for the cycles requiring the greatest correction

A shift in position of 1 bucket in the Main Injector is of little consequence, so the extraction can be synchronized to the position of the notch in the Booster.

All of these cogged cycles were taken with the same baseline trajectory. As time passes, the Booster magnet frequency drifts resulting in greater relative slippage. We anticipate being able to take a baseline on the first cycle of every multibatch cycle. This will result in significantly smaller variation and smaller radial displacements.

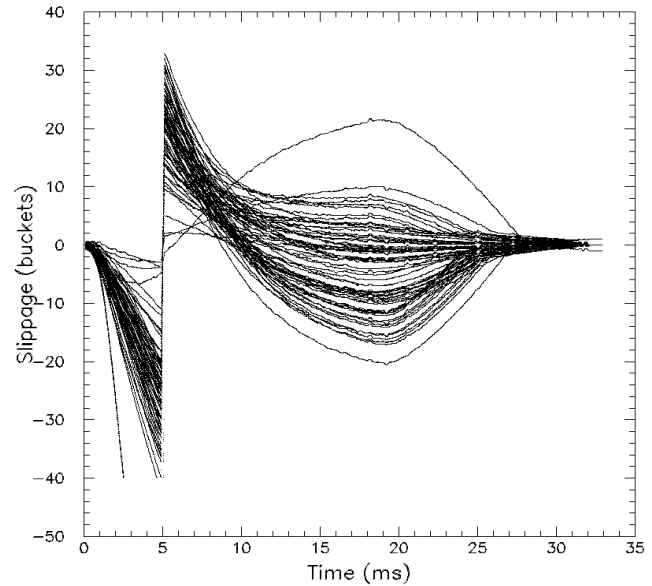

Figure 3: A series of successfully cogged cycles. About 90% achieve an error of zero buckets; the other 10% are at ± 1 bucket.

## CONCLUSION

We have analyzed the sources of cycle-to-cycle variation in the Booster beam's longitudinal trajectory. Avoidable sources of variation were eliminated and a system designed to correct for the remaining variation. This system measures variation and applies fast radial feedback to compensate, allowing multi-batch operation of the Main Injector without unacceptable Booster losses.